# Scalable $MoS_2$/Si Photodiode Arrays from Roll-to-Roll Mechanical Exfoliation

*Yigit Sozen[1], Thomas Pucher[1], Andres Castellanos-Gomez[1]*

*[1]2D Foundry research group. Instituto de Ciencia de Materiales de Madrid (ICMM-CSIC), Madrid, E-28049, Spain.*

*corresponding authors yigit.sozen@csic.es, andres.castellanos@csic.es

## ABSTRACT

We present a scalable, all-dry route to $MoS_2$/Si photodiodes based on high-throughput roll-to-roll mechanical exfoliation. The method yields dense films of large-lateral-size $MoS_2$ nanosheets and enables the formation of heterojunctions that operate in the photovoltaic regime when transferred onto locally etched Si regions. Individual devices show robust, fast photoresponses and broadband wavelength sensitivity. We further implemented a linear pixel array demonstrating position sensing capability and a 4×4 array as a proof-of-concept imaging sensor. These results indicate that roll-to-roll mechanical exfoliation can overcome common limitations of solution-processed films to enable low-cost, self-powered optoelectronic detectors for sensing and imaging applications.

INTRODUCTION

Two-dimensional (2D) semiconductors, and in particular transition metal dichalcogenides such as molybdenum disulfide ($MoS_2$), have attracted significant attention because of their thickness-tunable bandgaps, strong light–matter interaction and potential for integration in flexible and miniaturized optoelectronic systems.[1,2]

Beyond their fundamental properties, 2D semiconductors have been actively explored as building blocks for photovoltaic devices (i.e. PN junctions).[3–7] Proof-of-concept solar cells and photodiodes based on single flakes or stacked heterostructures have demonstrated that atomically thin materials can harvest light and generate photovoltages, with the advantages of mechanical flexibility, tunable bandgaps, and compatibility with transparent and flexible substrates. [4,5,8–10] Beyond the combination of 2D materials, there is a growing interest in mixed dimensionality architectures, [11] where a 2D material is interfaced with a bulk semiconductor to form a heterojunction exploiting all the advantages of bulk semiconducting doping control and micro-fabrication capabilities. In particular, combining 2D semiconductors with established silicon technology is especially attractive, since 2D/Si heterojunctions can provide a built-in electric field for efficient carrier separation and can be readily integrated with CMOS-compatible substrates. [12,13]

Despite of the interest on mixed dimensionality combining 2D semiconductors with silicon, the limited lateral size of exfoliated flakes hampers a proper scaling up and batch fabrication. Large-area and high-quality $MoS_2$ films are typically produced by chemical vapor deposition (CVD) or epitaxial growth methods, which, despite yielding superior electrical properties, are limited by high processing temperatures, complexity, and cost and they require non-trivial transfer processing of the as-grown films. [14–16] Alternatively, liquid-phase exfoliation combined with inkjet or spray printing offers attractive low-cost and scalability, [17–19] yet the small flake size and weak inter-flake coupling in such films commonly produce high junction resistance that limits photovoltaic performance on Si substrates. [20–23]

To overcome this bottleneck we employ an all-dry roll-to-roll mechanical exfoliation[24,25] process that yields dense, large-lateral-size $MoS_2$ films (2–8 μm lateral size, 10–60 nm thickness) and promotes enhanced nanosheet-to-nanosheet conduction [26] and intimate contact with freshly etched silicon. In the following we present the fabrication of $MoS_2$/Si heterojunction photodiodes using this approach and their comprehensive characterization: evaluation of photovoltaic current-voltage (*IV*) curves and time-resolved response on single devices, demonstration of spatially resolved detection with a linear array, and a proof-of-concept 4×4 pixel array that detects projected images.

## RESULTS AND DISCUSSION

### *Photodiode fabrication.*

Device fabrication is summarized in Figure 1a. First, electrode patterns were defined by maskless photolithography on $SiO_2$/Si substrate, followed by Cr/Au (≈5/50 nm) deposition via thermal evaporation, and subsequent lift-off (panels a.1 to a.4). Afterwards windows for junction formation were defined on the $SiO_2$ layer by a second aligned lithography step followed, and selectively etched with Armour Etch cream to expose the underlying highly p-doped Si layer (panels a.5 to a.7). [27] Lastly, the dry transfer of a dense film of n-type $MoS_2$ flakes, produced by roll-to-roll high-throughput mechanical exfoliation, onto the exposed Si regions (via thermal release at 110 °C for 5 min) forms the $MoS_2$/Si PN junction interface while simultaneously bridging the Si layer to the Au pads, thereby establishing the electrical pathway required for photodiode operation (Figure 1.a.8). We address the reader to the Materials and Methods section for more fabrication details.

The high doping concentration of the Si substrate affects the device behavior in several ways. First, the high acceptor concentration reduces the depletion width on the Si side of the $MoS_2$/Si heterojunction, causing the built-in electric field to be localized mainly near the $MoS_2$/Si interface. Since the photovoltaic response originates from the separation of photogenerated carriers by this interfacial built-in electric field, the device photovoltaic performance is expected to be strongly influenced by the $MoS_2$/Si interfacial quality, including defects, trap states, and interfacial recombination processes. In addition, the highly doped Si substrate serves not only as the p-type side of the $MoS_2$/Si heterojunction but also as a conductive contact in the two-probe measurement configuration owing to its low resistivity. This reduces the contribution of the Si substrate resistance to the measured electrical characteristics, allowing the measured response to more directly reflect the $MoS_2$/Si heterojunction and the transferred $MoS_2$ flake network.

To investigate the spatially resolved photodetection capability of the PN junctions, we designed two distinct device architectures. A linear photodiode array, in which the devices are arranged in one dimension, was used to confirm position-sensitive photodetection, while a 4×4 pixelated photodiode array was used to demonstrate the potential of the platform for image-sensing applications. Figure 1b shows the linear array structure after transfer of the $MoS_2$ film, consisting of a long rectangular Si opening with Au pads arranged along one side. During fabrication, the spacing between the Au pads and the Si opening was minimized to facilitate bridging by the transferred $MoS_2$ flakes. Two or three transfer steps were implemented depending on the degree of $MoS_2$-mediated bridging achieved between the Si opening and Au pads. Figure S1 shows a representative portion of the pixel channel area covered with $MoS_2$ flakes. Figure 1c shows the 4×4 device layout used for subsequent image-sensing measurements, in which four Si openings are vertically arranged, with each opening surrounded by four Au pads.

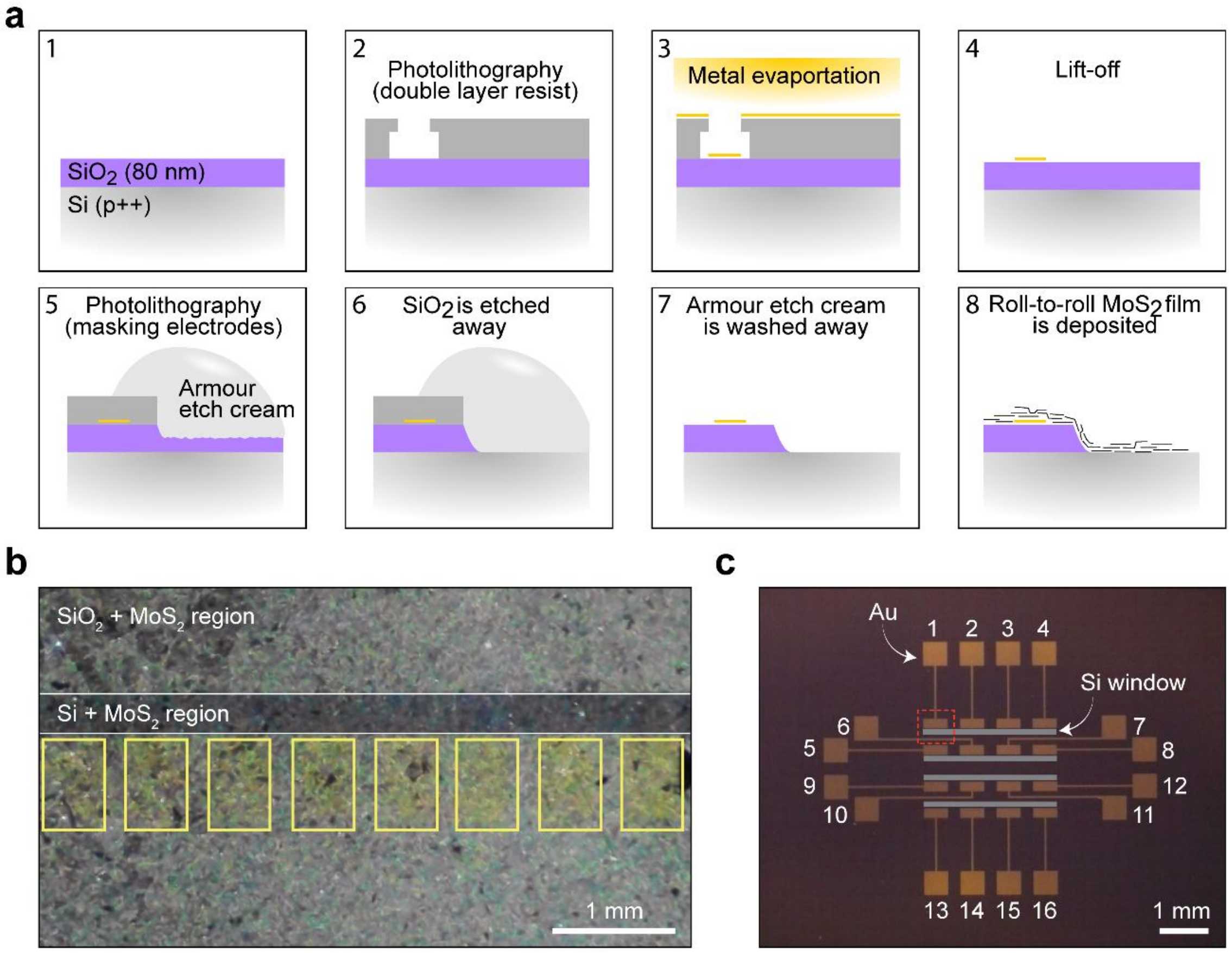

**Figure 1. Device fabrication.** (a) Schematic illustrates the main fabrication steps for $MoS_2$/Si heterojunction: (1) $SiO_2$/Si substrate (P-type Si, 80 nm thermal oxide); (2) first photolithography to define electrodes; (3) Cr/Au metal deposition; (4) lift-off, revealing the metal electrodes; (5) second aligned photolithography to define oxide windows; (6) selective $SiO_2$ removal with Armour Etch cream; (7) rinse/dry; (8) dry transfer of roll-to-roll $MoS_2$ film onto the exposed Si. (b) Optical image of a transferred $MoS_2$ film forming a linear array of photodetectors. Areas surrounded with yellow rectangles locates gold pads. (c) Photograph of completed 4×4 arrays of photodetectors chips just prior deposition of $MoS_2$. The etched silicon windows are illustrated schematically.

*Photodiode Characterization.*

Individual $MoS_2$/Si photodiodes were characterized in a custom-built probe station, operating in room conditions and equipped with a multimode optical fiber to deliver illumination from fiber coupled light sources (see Materials and Methods for details). The fiber's output was projected onto the device using a zoom lens, ensuring localized illumination of the active area with a homogeneous illumination power density. [28,29] Electrical measurements were performed using a Keithley 2450 Source Measure Unit (SMU).

Figure 2a shows the current-voltage (*IV*) characteristics of one of the $MoS_2$/Si photodiodes measured in the dark state and under illumination with a 660 nm LED. The illumination power is 0.48 mW with a 0.9 mm diameter spot size, fully covering the active region of the PN junction. Dark *IV* curve shows a marked rectifying behavior. Upon illumination, photogenerated current contributes to the dark current, in both forward and reverse regions of the *IV* curve.

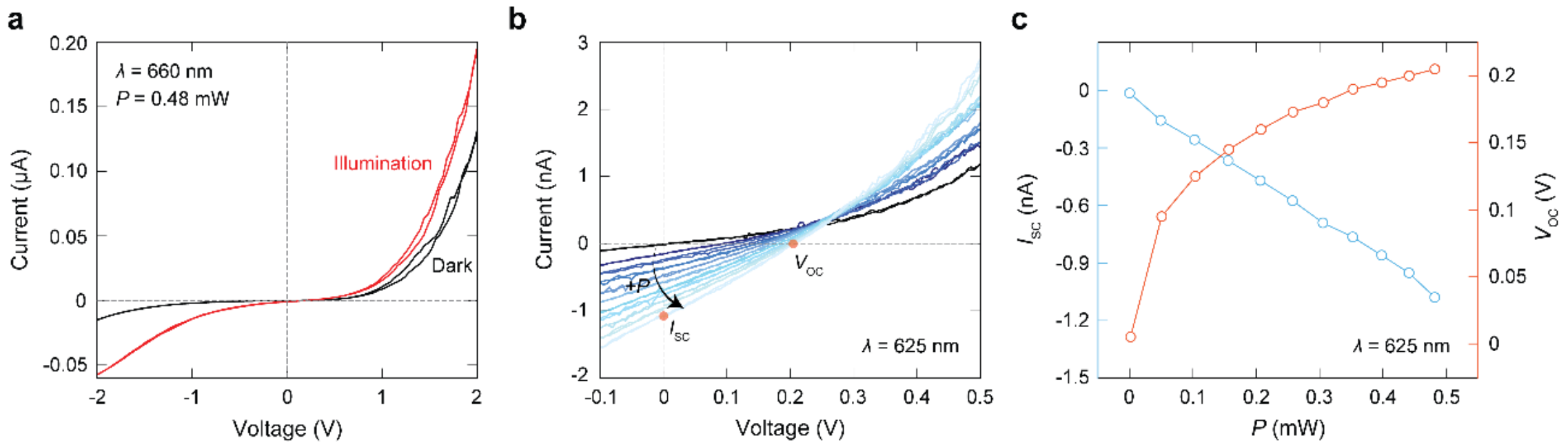

**Figure 2. Optoelectronic characterization of a $MoS_2$/Si photodiode.** (a) Current–voltage (*IV*) characteristics of a representative $MoS_2$/Si photodiode measured in the dark (black) and upon illumination (red). (b) Power-dependent *IV* characteristics under illumination at $\lambda = 625$ nm. The black trace is the dark *IV*; blue traces correspond to increasing incident optical power (lighter color = higher power). Under increasing illumination power the device shows a clear photovoltaic shift, with growing short-circuit current ($I_{sc}$) and open-circuit voltage ($V_{oc}$). (c) Extracted short-circuit current (blue markers, left axis) and open-circuit voltage (orange markers, right axis) as a function of incident optical power.

Figure 2b shows the evolution of *IV* traces of a single $MoS_2$/Si photodiode with increasing illumination power at $\lambda = 625$ nm. These *IV*s have been focused in a smaller voltage range to better resolve the photovoltaic generated photocurrent features. The black curve corresponds to the dark condition; progressively lighter blue traces correspond to increasing incident power. Under illumination the junction develops a clear photovoltaic response: the $IV$ curve shifts such that a finite short-circuit photocurrent ($I_{sc}$) appears at $V = 0$ and an open-circuit voltage ($V_{oc}$) develops where the net current crosses zero. Identical characterizations were performed on other four different $MoS_2$/Si photodiodes under the same experimental conditions to obtain device-to-device statistics (see Supporting Information, Figure S2). All devices exhibited clear photovoltaic responses with similar trends in $V_{oc}$ and $I_{sc}$ as a function of incident light power, indicating reproducible $MoS_2$/Si heterojunction formation.

To quantify these trends we extract $I_{sc}$ and $V_{oc}$ from the measured traces and plot them as a function of incident light power as represented in Figure 2c. The $I_{sc}$ increases approximately linearly with illumination power over the measured range. The $V_{oc}$ shows a sublinear increase with power that is well described by the diode equation dependence:

$$V_{OC} \approx \frac{nk_BT}{q} ln\left(\frac{I_{ph}}{I_S} + 1\right),$$

with $n$ the ideality factor, $k_B$ the Boltzmann's constant, $T$ the temperature, $q$ the elemental charge and $I_{ph}$ and $I_s$ being the photocurrent and the saturation current respectively. Because of the linear dependence of the generated photocurrent with power, $V_{oc}$ is thus expected to grow proportionally to ln($P$). The observed trends, *i.e.* linear scaling of $I_{sc}$ and logarithmic-like increase of $V_{oc}$, confirm that the devices operate in the photovoltaic regime and that charge separation at the $MoS_2$/Si heterojunction dominates the optoelectronic response. The plotting $V_{oc}$ as a function of ln($P$) yields a linear relationship whose slope ($S$) equals $nk_BT/q$. Therefore, the ideality factor ($n$) can be obtained from $n = \frac{q}{k_B T} S$. Using the room-temperature thermal voltage ( $k_B T/q \approx$ 25.9 mV) and the slope $S$ extracted from the linear fit of $V_{oc}$ as a function of ln($P$) (see Supporting Information, Figure S3), the ideality factor was determined to be 1.9, indicating that recombination processes playing a significant role in carrier transport across the junction. The extracted ideality factors for additional pixels are presented in Figure S4.

The observation of a clear photovoltaic response under zero external bias distinguishes these devices from many previous $MoS_2$/Si heterojunction implementations. For devices based on liquid-phase exfoliated $MoS_2$ films drop-casted or inkjet printed onto Si, photovoltaic characteristics are generally weak or absent, and external bias is required to extract photocurrent due to the high flake-to-flake resistance that limits carrier transport across the film. [30]

Devices based on CVD $MoS_2$ flakes transferred onto Si have demonstrated high responsivity and even avalanche multiplication at large reverse bias, but their operation typically relies on externally applied voltage and does not yield sizeable $V_{oc}$ under illumination. [31] Also, large-area $MoS_2$/Si photodiodes formed by sulfurizing sputtered $MoO_3$ on Si exhibit robust photoconductive gain and fast temporal response, yet are operated in the photoconductive regime at finite bias and did not report zero-bias photovoltaic characteristics. [32]

In our devices, the $MoS_2$ films are transferred via an entirely dry process, which minimizes interfacial contamination and promotes a direct van der Waals contact between the $MoS_2$ film and the exposed Si surface. This clean interface facilitates the formation of a built-in electric field at the 2D/Si junction, allowing the devices to operate in the photovoltaic mode and to generate sizeable open-circuit voltages ($V_{oc} \approx 0.2$ V, see also Supporting Information datasets) under illumination without external bias. The observation of a measurable $V_{oc}$ indicates that the $MoS_2$ is not acting merely as a passive absorber layer but electrically participates in carrier separation at the heterojunction, distinguishing this behavior from that of solution-processed $MoS_2$ films on Si, where interfacial contamination and native oxide often suppress photovoltaic response and necessitate biased photoconductive operation, or even CVD $MoS_2$ films where the wet assisted transfer process could also lead to interfacial contamination.

*Spatially Resolved Photodetection with a Linear Array.*

To explore the capability of the devices as spatially sensitive detectors, we fabricated a 16-pixel linear photodiode array. A highly focused light spot ($\lambda = 625$ nm) was projected onto the array such that only one pixel and its immediate neighbors were illuminated. The photocurrent response of all pixels was recorded and visualized as a per-pixel heat map of the $V_{oc}$ (see Figure 3). The map shows a pronounced peak at pixel 12 ($V_{oc} \approx 0.17$ V), with smaller yet measurable responses at adjacent pixels (pixel 11 $\approx$ 0.028 V, pixel 13 $\approx$ 0.098 V), and baseline values near zero (within the experimental error) elsewhere. This spatial profile reproduces the position of the optical spot, indicating that the array can function as a one-dimensional position-sensitive detector in the photovoltaic mode.

Although position-sensitive photodetection has been explored using 2D materials in photoconductive or phototransistor geometries, such implementations require external bias to generate sufficient signal and often rely on photoconductive gain or bolometric mechanisms. [33] In contrast, the $MoS_2$/Si junctions used here operate entirely without bias and rely on photovoltaic carrier separation at the 2D/Si interface, eliminating dark-power consumption and simplifying readout. These device-level distinctions are relevant for low-power imaging or sensing architectures, where voltage-mode output and self-powered operation are advantageous.

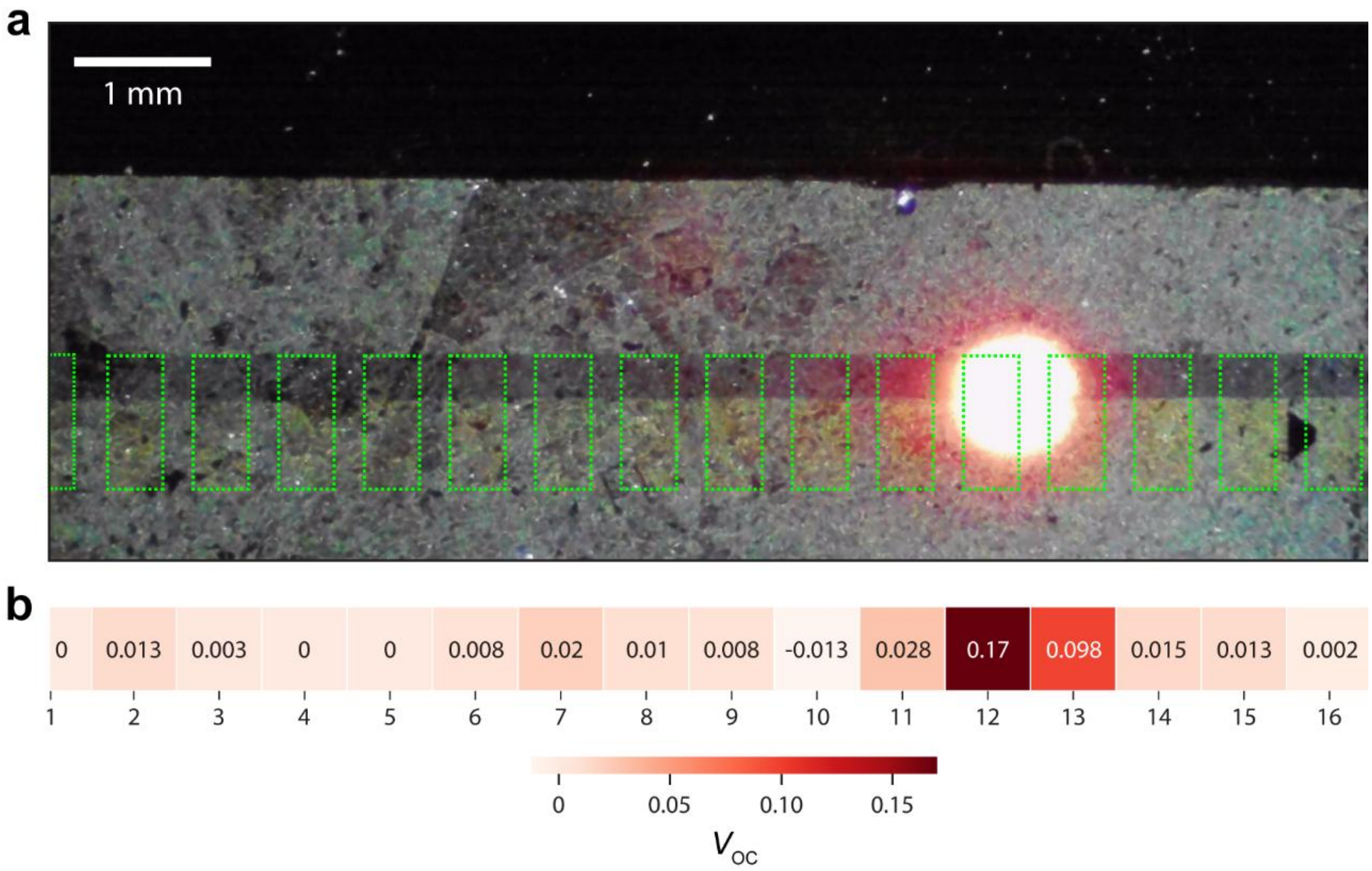

**Figure 3. Linear $MoS_2$/Si photodiode array used as a position-sensitive light sensor.** (a) Optical micrograph of the array under focused illumination; dashed green rectangles indicate the individual pixel areas (gold pad and PN junction) used for electrical readout. (b) Heat-map representation of the open-circuit voltage ($V_{oc}$) measured at each pixel when the focused spot is projected onto the array. The largest $V_{oc}$ signals are localized at pixels 11–13, closely matching with the position of the optical spot.

*Image Sensing Demonstrated with a 4×4 Array.*

In a final experiment, we extended the concept to a 4×4 photodiode array to demonstrate imaging capabilities. An aluminum shadow mask was fabricated using a laser engraving system to create a user-defined pattern.[34] This mask was used to project an image onto the array by illuminating it with a white LED source through a plano-convex lens. The short-circuit current, $I_{sc}$, generated by each pixel was measured and compiled into a heat map (see Figure 4). The reconstructed heat map corresponded closely with the projected image, validating the feasibility of using such arrays for rudimentary image sensing.

Figure S5 shows the extracted $V_{OC}$ and $I_{SC}$ values for each pixel in the 4×4 photodiode array. Although all pixels exhibit photovoltaic behavior, pixel-to-pixel variations are observable. The variation is more pronounced in $I_{SC}$ than in $V_{OC}$, which is consistent with the logarithmic dependence of $V_{OC}$ on photocurrent. These variations are expected for devices composed of flake networks rather than individual flakes, because randomly assembled flakes with different lateral sizes, thicknesses, and overlap areas create distinct percolation pathways across the device.[22,35] This network nonuniformity can lead to pixel-dependent differences in effective series resistance, charge transport, and local $MoS_2$/Si junction characteristics.

Furthermore, the $MoS_2$ flake thickness can influence the photodiode performance through its effect on optical absorption and carrier collection. In photodiodes, the useful photocurrent is mainly generated by photocarriers formed within, or sufficiently close to, the depletion region of the junction, where they can be efficiently collected by the built-in electric field before recombination. A previous study reported that optimum device performance can be achieved when the $MoS_2$ thickness is comparable to the $MoS_2$ depletion-layer width, since this condition provides a favorable balance between optical absorption and carrier collection.[36] In our devices, however, the $MoS_2$/Si interface is formed by randomly transferred $MoS_2$ flakes with nonuniform thicknesses, varying mostly within the range of 20–40 nm per transfer, as reported in our previous studies.[24,25] Therefore, the relative thickness of the $MoS_2$ flakes with respect to the local depletion/electric-field region may vary across different $MoS_2$/Si junctions. This thickness variation, together with differences in flake-network structure, flake connectivity, and interfacial contact area, can contribute to the observed pixel-to-pixel variations in the electrical and photovoltaic characteristics.

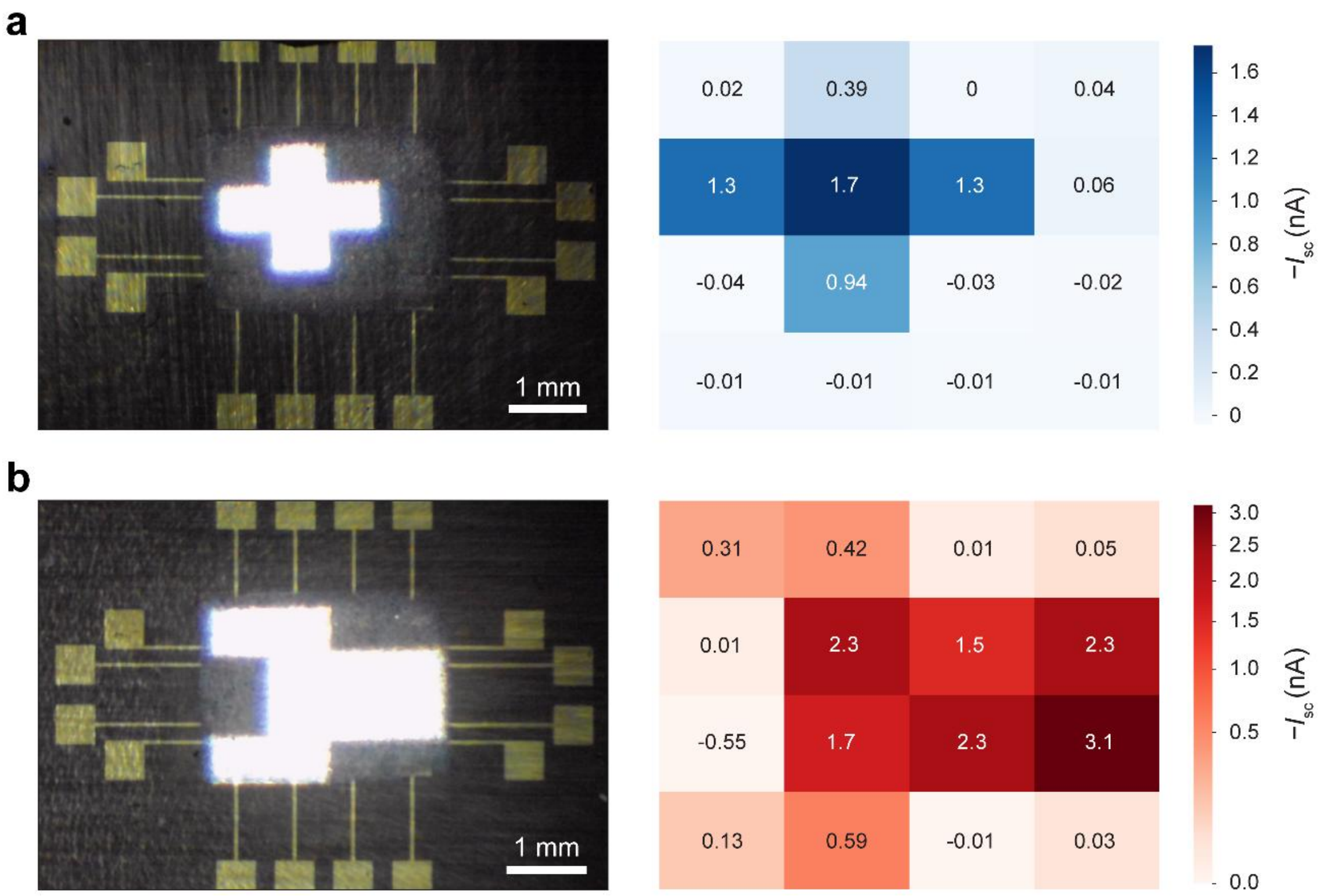

**Figure 4. Proof-of-concept imaging with a 4×4 MoS2/Si photodiode array.** (a) and (b) Optical image of the array under two distinct patterned illuminations (left) and resulting 4×4 heat map extracted from the measured short-circuit current ($I_{sc}$) of each pixel (right). Laser-engraved aluminum masks were used to project the patterns onto the pixel region. Note that the white light pattern is saturated as long acquisition time imaging was necessary to capture the pattern and the rest of the un-illuminated device.

The demonstration of a 4×4 $MoS_2$/Si array operating in photovoltaic mode enables rudimentary image reconstruction without the need for external bias. This is in contrast to most previously reported $MoS_2$ photodetector arrays, operating in the photoconductive regime and rely on external bias to extract photocurrent. Inkjet-printed $MoS_2$ nanosheet arrays, for example, have recently achieved image readout with responsivities exceeding 500 A $W^{-1}$, but require several volts of applied bias. [37,38] CVD-grown $MoS_2$ arrays have demonstrated fast response times, yet operate as back-to-back Schottky photoconductors with negligible photovoltaic behavior and no imaging has been demonstrated. [39] Recent deterministic assembly of (manually) mechanically exfoliated single flakes has also enabled high-density $MoS_2$ imaging arrays. [40] In those devices, nonetheless, pixel-to-pixel variability in responsivity and dark current is corrected through pixel-wise calibration during post-processing, allowing the reconstructed image to reflect the illumination pattern rather than device non-uniformity.

In our devices, the imaging functionality arises from the built-in field at the $MoS_2$/Si heterojunction, allowing patterned illumination to be transduced directly into pixel-resolved short-circuit currents ($I_{sc}$) without bias. While the present spatial resolution is modest, limited primarily by pixel pitch and optical projection, the ability to perform zero-bias image readout with a scalable and low-temperature fabrication route highlights a complementary pathway toward low-power 2D-material-based imagers. Although pixel-to-pixel variations are present, the reconstructed image quality was sufficient without pixel-wise calibration.

## CONCLUSIONS

We present a simple, scalable method to make $MoS_2$/Si photodiodes by transferring large-area films produced with roll-to-roll mechanical exfoliation. The process is all-dry, compatible with standard microfabrication, and yields continuous $MoS_2$ films that form functioning photovoltaic junctions with silicon. Individual devices operate clearly in the photovoltaic regime. Photogenerated electron–hole pairs are separated by the built-in field at the $MoS_2$/Si heterojunction and can be extracted as a measurable short-circuit photocurrent ($I_{sc}$) even with the device unbiased; when the circuit is open the same mechanism produces a finite open-circuit voltage ($V_{oc}$). We further showcase the potential of this batch fabrication process by fabricating a 16 pixels linear array and a 4×4 pixel matrix array demonstrating light sensitive position detection and imaging capability. These proof-of-concept results indicate that mechanically exfoliated 2D films can be used to build low-cost, self-powered light sensors and simple imaging arrays. With further improvements in contacts, pixel density and readout, this approach could be extended toward higher-performance, inexpensive optoelectronic systems.

MATERIALS AND METHODS

## Device fabrication

*Substrate preparation and electrode patterning.* Photodiodes were fabricated on commercially available Si wafers coated with an 80 nm thermally grown $SiO_2$ layer. The Si substrate was he avily p-doped, with a resistivity of 0.001–0.005 Ω·cm, corresponding to a dopant concentration on the order of $10^{19}$–$10^{20}$ $cm^{-3}$. Electrode patterns were defined via maskless photolithography using a SmartPrint system (Microlight 3D SmartPrint UV). A two-layer lithography process, using LOR2A and AZ1505, allows to define a resist undercut to achieve straight (‘doggy ears’ free) metal edges. After resist exposure and development, a bilayer of chromium (~5 nm) and gold (~45 nm) was deposited by thermal evaporation. A subsequent lift-off process in DMSO produced well-defined Cr/Au electrode arrays.

*Window definition and $SiO_2$ etching.* A second aligned photolithography step with a simpler one-layer process (AZ1512HS), defined windows in the resist at the locations intended for active junction formation, right next to the pre-existing metal electrodes. The exposed $SiO_2$ (not masked by the resist) was then etched using Armour Etch glass etching cream, as the photoresist protected the rest of the substrate. The etching process is self-limiting, stopping when reaching the underlying silicon. Following etching, the substrates were thoroughly rinsed with Milli-Q water and dried under a nitrogen stream. The remaining photoresist was removed in acetone, followed by rinsing with isopropanol and blow-drying with nitrogen.

*$MoS_2$ film deposition via roll-to-roll mechanical exfoliation.* Immediately after oxide removal, a dense film of $MoS_2$ nanosheets was transferred onto the substrate via a roll-to-roll mechanical exfoliation technique, as described by Sozen et al. [24]. In this method, two polyoxymethylene (POM) cylinders (with incommensurate perimeters selected at a 53:23 ratio) are uniformly coated with double-sided adhesive tape (Nitto SPV 224). When the cylinders are pressed together under a controlled force (20–40 N) and rotated at approximately 1500 rpm for 20 seconds, shear forces at the interface induce massive parallel exfoliation of a bulk $MoS_2$ crystal. The resulting tape is uniformly loaded with nanosheets exhibiting lateral dimensions of 2–8 μm and thicknesses from 10 to 60 nm. For transfer, the tape is brought into contact with the etched $SiO_2$/Si substrate and heated at 110 °C for 5 minutes, ensuring a high-yield, dry deposition of the $MoS_2$ film with the basal planes parallel to the substrate. This process achieves an area coverage of up to approximately 75%, thereby facilitating efficient formation of $MoS_2$/Si heterojunctions.

**Electronic and optical measurements.**

All electrical and optical measurements were carried out at room temperature on a home-built probe station. Dark and illuminated *IV* characteristics were recorded using a Keithley 2450 source meter unit. For power-dependent photocurrent experiments we illuminated devices with a fiber-coupled 625 nm LED (Thorlabs M625F2); the fiber output was projected onto the sample through the probe-station zoom lens optics. For the imaging demonstrations, patterned illumination was produced by a fiber-coupled white LED (model MCWHF2) and an aluminum shadow mask fabricated with a laser engraving system (settings: 100% power, 1 mm $s^{-1}$); the pattern was focused onto the device using a bi-convex lens (Thorlabs LB1471). Incident optical power was measured with a Thorlabs PM100D optical power meter together with an S120C photodiode power sensor.

## Use of generative AI tools

The authors used ChatGPT (OpenAI) solely to assist with language editing and readability.[41] All scientific ideas, experiments, data analysis, and interpretations are the authors' own. The final text was reviewed and edited by the authors, who take full responsibility for its content.

## ASSOCIATED CONTENT

### Supporting Information

Supplementary Information includes:

**Figure S1.** Representative channel area of a pixel covered with $MoS_2$ flakes.

**Figure S2.** Optoelectronic characterization of different $MoS_2$/Si photodiodes.

**Figure S3.** Extraction of the diode ideality factor.

**Figure S4.** Distribution of ideality factors across $MoS_2$/Si photodiodes.

**Figure S5.** Photovoltaic parameters from the 4×4 $MoS_2$/Si photodiode array.

## AUTHOR INFORMATION

### Corresponding Authors

Yigit Sozen yigit.sozen@csic.es, Andres Castellanos-Gomez andres.castellanos@csic.es

### Author Contributions

Y.S. led data curation, formal analysis, investigation, and device fabrication methodology, and contributed to writing the original draft. T.P. led the methodology for substrate preparation and electrode patterning using photolithography and contributed to data curation, formal analysis, investigation, and writing of the original draft. A.C.-G. led conceptualization, funding acquisition, project administration, resources, supervision, and writing of the original draft, and contributed to the methodology. All authors contributed equally to review and editing of the manuscript.

## ACKNOWLEDGEMENTS

We acknowledge the Severo Ochoa Centers of Excellence Program through the grant CEX2024-001445-S funded by MICIU/AEI/10.13039/501100011033. A.C-G. acknowledges support from Grant PDC2023-145920-I00 funded by MICIU/AEI/10.13039/501100011033 and by the European Union NextGenerationEU/PRTR, and from Grant PID2023-151946OB-I00 funded by MICIU/AEI/10.13039/501100011033 and by ERDF/EU. A.C.-G. also acknowledge funding from the European Research Council (ERC) through the ERC-2024-PoC StEnSo project (Grant Agreement No. 101185235) and the ERC-2024-SyG SKIN2DTRONICS project (Grant Agreement No. 101167218). Y.S. acknowledges support from Grant PRE2021-098348 funded by MICIU/AEI/10.13039/501100011033 and by ESF+.

# Supporting Information:

# Scalable $MoS_2$/Si Photodiode Arrays from Roll-to-Roll Mechanical Exfoliation

*Yigit Sozen[1], Thomas Pucher[1], Andres Castellanos-Gomez[1]*

*[1]2D Foundry research group. Instituto de Ciencia de Materiales de Madrid (ICMM-CSIC), Madrid, E-28049, Spain.*

*corresponding authors yigit.sozen@csic.es, andres.castellanos@csic.es

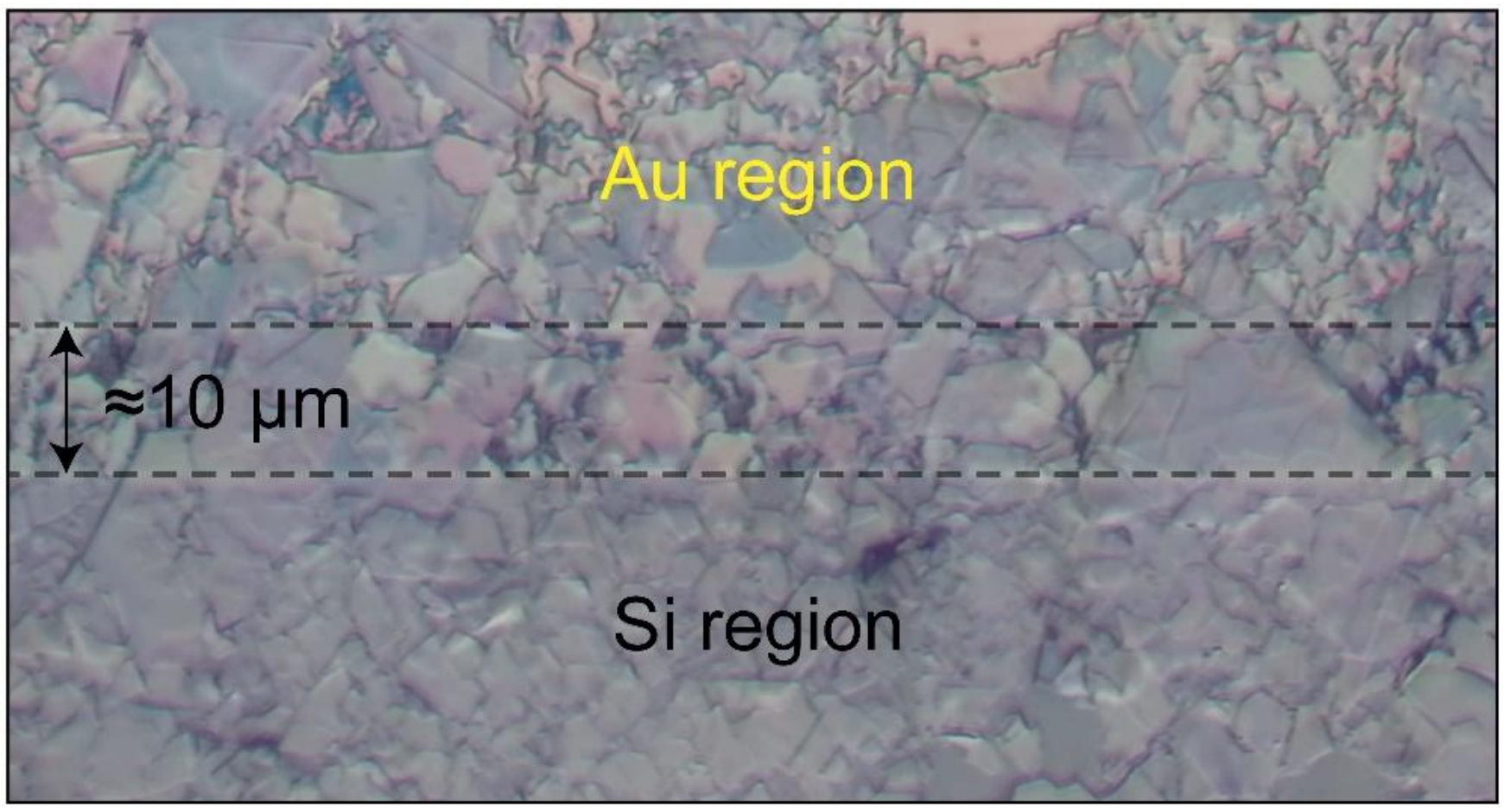

**Figure S1. Representative channel area of a pixel covered with $MoS_2$ flakes.** Optical microscope image showing a representative portion of the channel area of a pixel. The Au electrode region is located on the insulating $SiO_2$ layer and is laterally separated from the exposed Si window by an approximately 10 µm $SiO_2$-covered gap. The transferred $MoS_2$ flakes bridge this gap, providing electrical connection between the Au electrode and the exposed Si region, while the $MoS_2$/Si heterojunction is formed only at the exposed Si window.

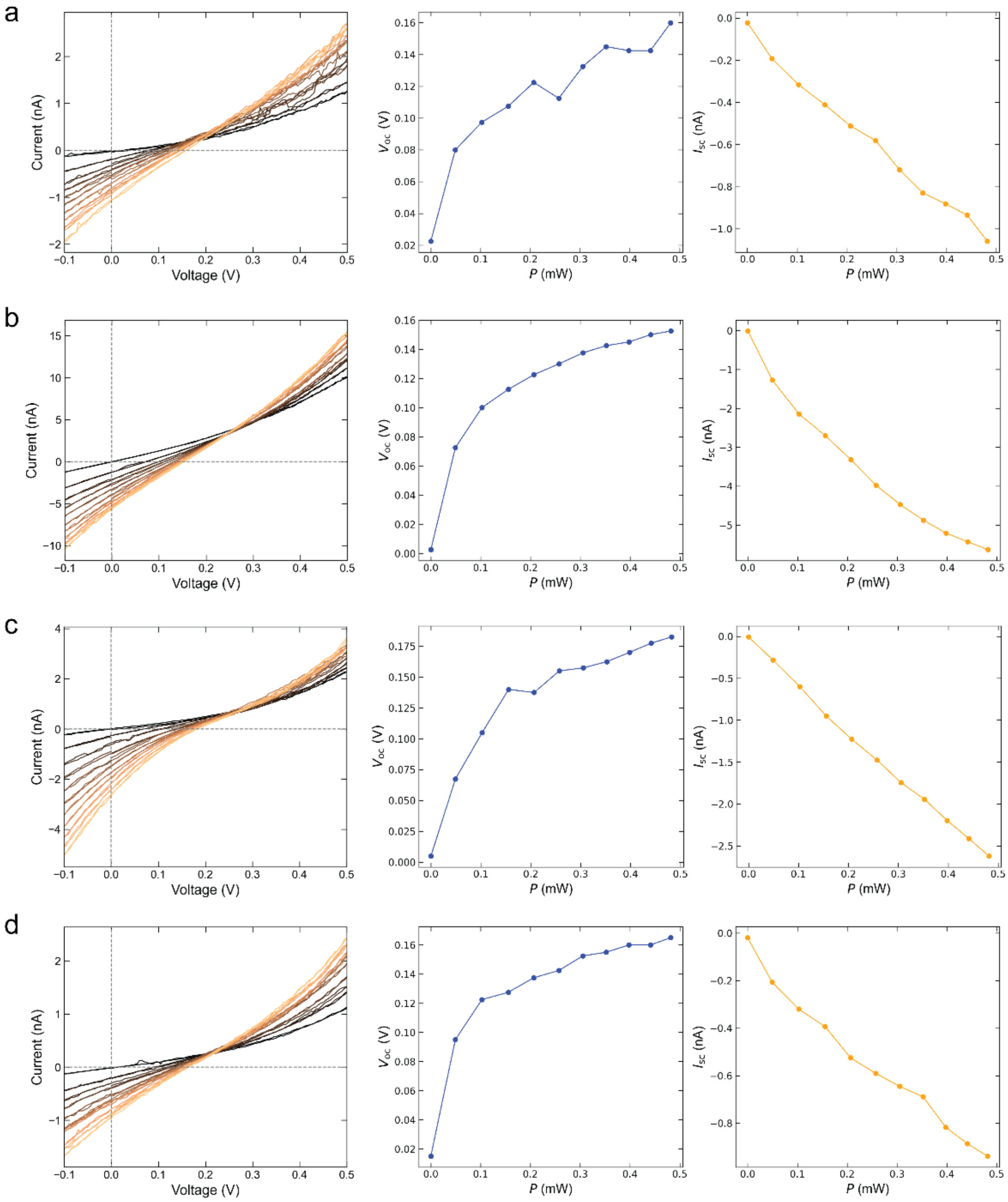

**Figure S2. Optoelectronic characterization of different $MoS_2$/Si photodiodes.** (a–d) Current–voltage (*IV*) characteristics measured under increasing illumination power (*P*), and the extracted open-circuit voltage ($V_{OC}$) and short-circuit current ($I_{SC}$) values plotted as a function of $P$, shown in the left, middle, and right panels, respectively.

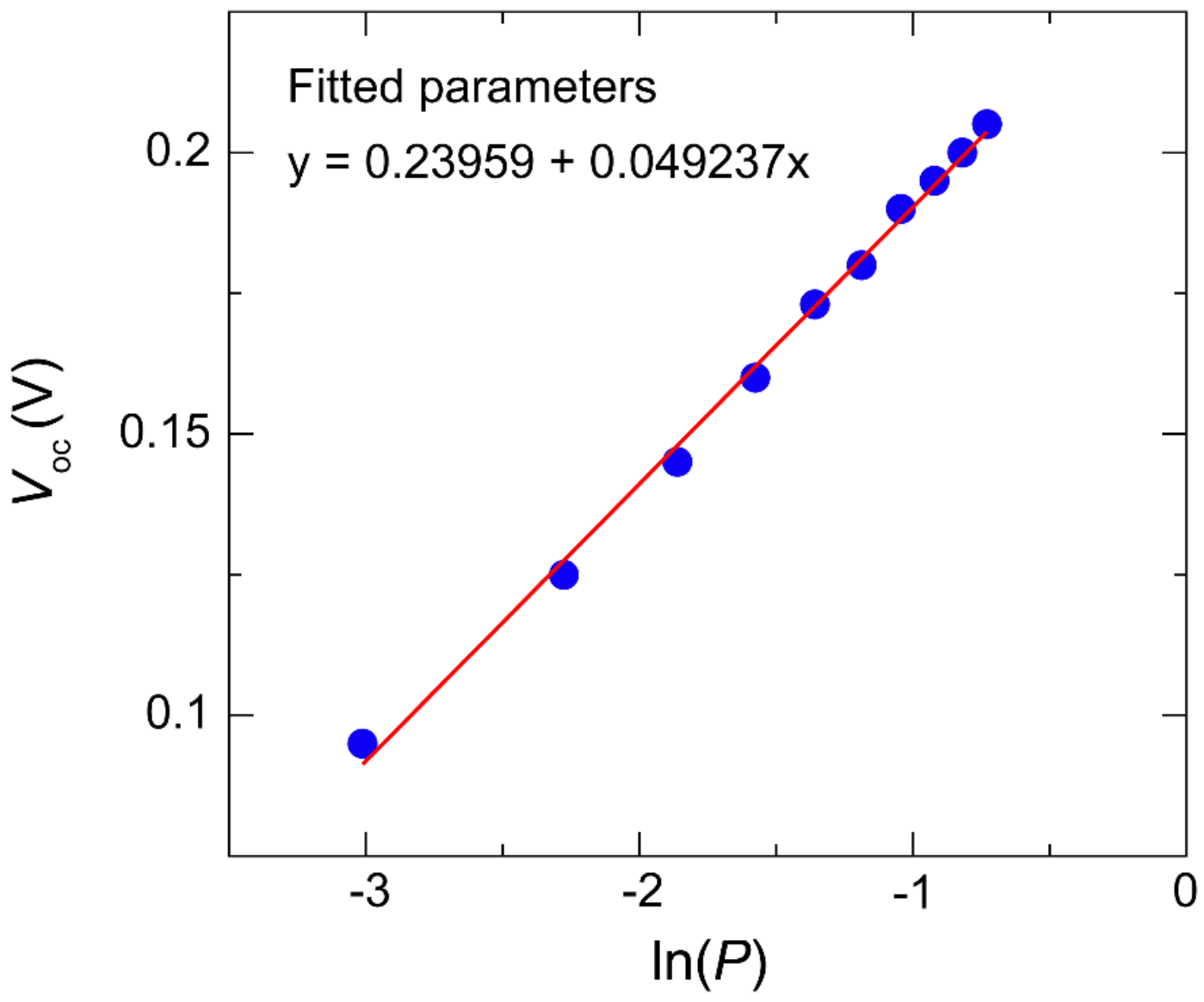

**Figure S3. Extraction of the diode ideality factor.** $V_{OC}$ versus $\ln(P)$ with corresponding linear fit (solid red line) used for extraction of the ideality factor.

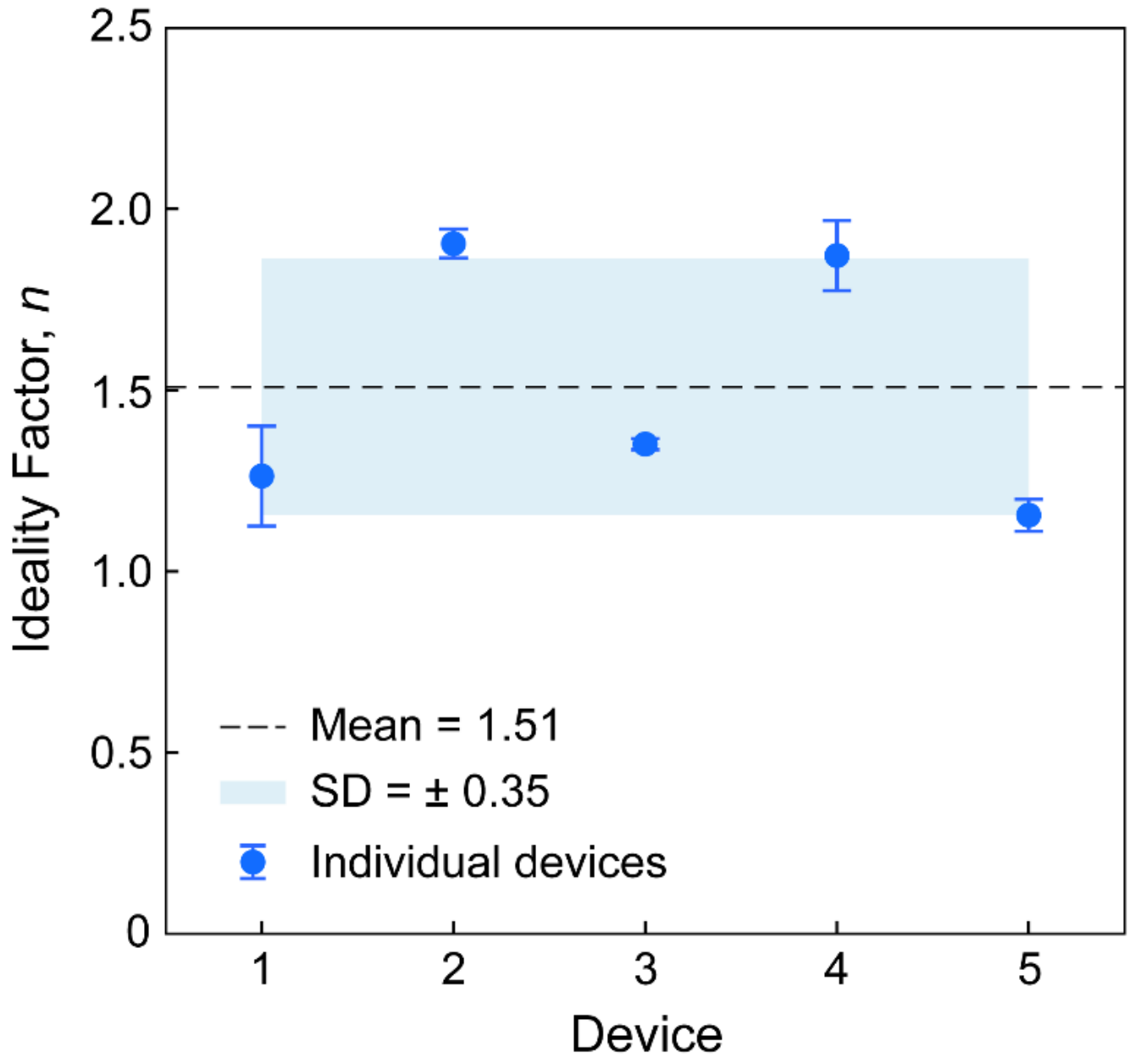

**Figure S4. Distribution of ideality factors across $MoS_2$/Si photodiodes.** Plot shows the ideality factors extracted from five $MoS_2$/Si pixels in linear pixel array. The data points represent the ideality factor values obtained from the linear fit of the $V_{OC}$ versus $\ln(P)$ characteristics for each pixel. The vertical error bars indicate the uncertainty of the extracted ideality factor, calculated from the standard error of the fitted slope. The dashed horizontal line represents the average ideality factor, $n_{\text{avg}} = 1.51$, while the shaded region corresponds to the device-to-device standard deviation, $n_{\text{avg}} \pm \text{SD}$, with $\text{SD} = 0.35$.

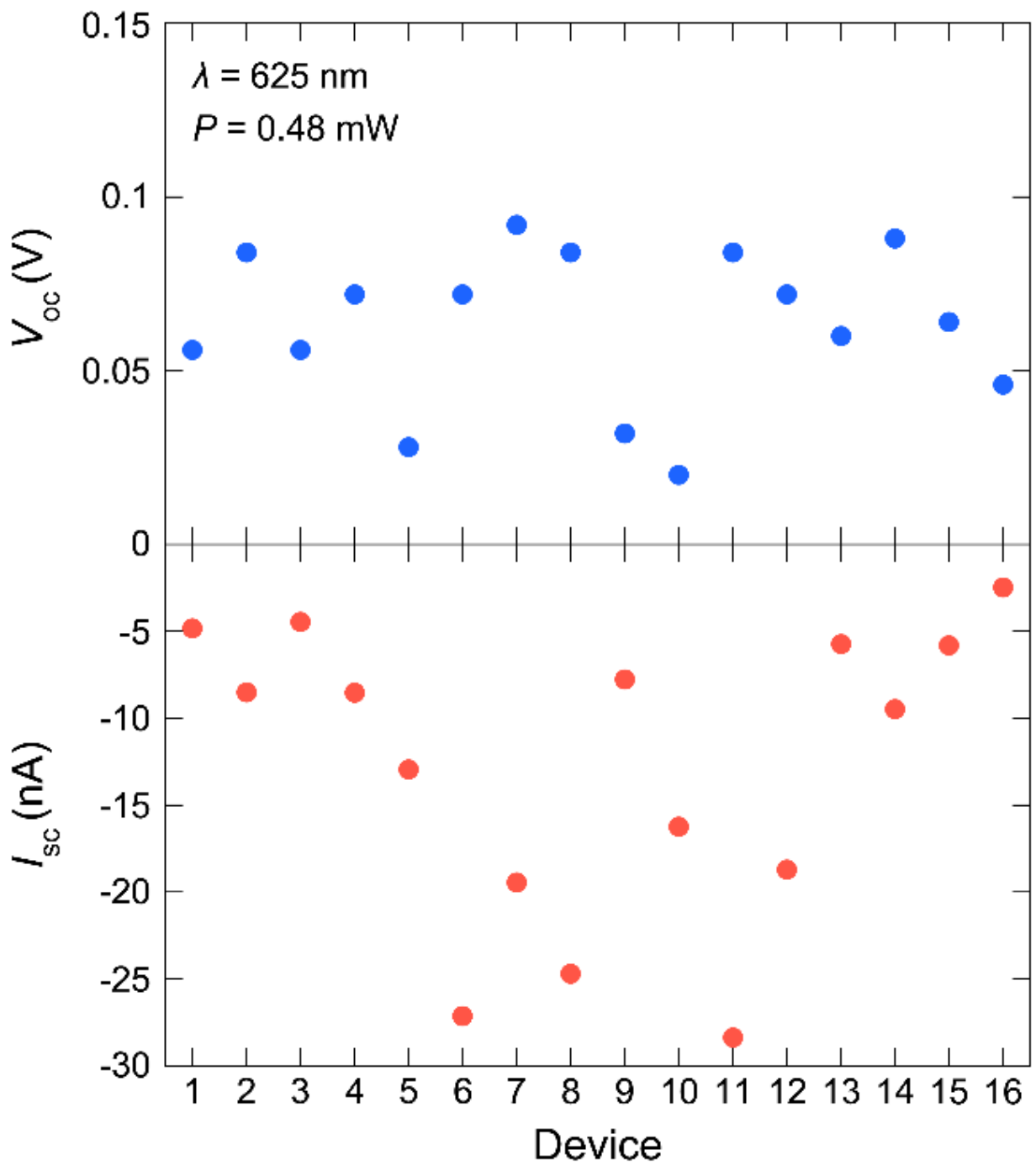

**Figure S5. Photovoltaic parameters from the 4×4 $MoS_2$/Si photodiode array.** The plot shows the extracted $V_{OC}$ and $I_{SC}$ values for all 16 pixels in the 4×4 photodiode array.